\documentclass[11pt,a4paper]{article}
\usepackage{graphicx}
\usepackage{amsmath}
\usepackage{amssymb}
\usepackage{epsfig}
\usepackage{minitoc}
\usepackage{authblk}
\usepackage{cite}
\usepackage{color}
\numberwithin{equation}{section}
\title{Kerr Black String Flow}

\author[a]{Meng Sun \thanks{sunmeg.89@gmail.com}}
\author[a,b]{Yong-Chang Huang \thanks{ychuang@bjut.edu.cn}}

\affil[a]{Institute of Theoretical Physics\\ Beijing University of Technology, 100124, Beijing, China}
\affil[b]{CCAST(WorldLab.), P.O. Box 8730, 100080, Beijing, China}
\date{}
\begin{document}
\maketitle
\begin{abstract}
	We give a general illumination about a rotating black string falling into a rotating horizon in dimension D=5.  It is a configuration about one smooth intersection between these two objects when the spacetime is axisymmetric and in the limit that the thickness of the black hole is much larger than the thickness of the black string. Following this configuration, we further extend them to the rotating and charged flows.

\end{abstract}

\flushbottom
\newpage

\section{Introduction}
The string theory is a very valuable tool for us to understand properties of the black hole.
One of the great successes in this field is the statistical explanation of the Bekenstein-Hawking entropy for the certain extremal and the near extremal black hole.
Different types of black holes have been studied well and all of them support the explanation for the Bekenstein-Hawking entropy since the first success in \cite{17}.
Recent works in this field have focused on the question that in stationary spacetime allowing the event horizon is not Killing horizon \cite{1}.
Some former works \cite{4,5,6} may forbid this construction.
However, from a physical perspective, there are two different surface gravities at the horizons where they connect two asymptotic spaces, which indicate different temperatures.
In thermodynamic theory, this is a description about a steady heat stream between two infinite heat reservoirs that preserve a fixed temperature gradient when the time is under evolution \cite{1}.

In article \cite{1}, a flowing horizon was constructed by a very impressive way. Basically, they build up the construction with a Schwarzschlid metric that is a static black hole solution.
However, in this article we extend the former solution to more general conditions.
We consider a Kerr metric to construct the flow, and it is easy to extend to a Kerr-Newman metric condition.
Even though the descriptions of horizon flows motivated by AdS/CFT \cite{2,18,19,20,21,22,Huang:2010yg,Huang:2,Fang-Fang:2013rra} have been studied well, as the same reason in \cite{1}, in our construction we do not need a negative cosmological constant to hold the system steadily.

We begin our construction from such a picture: a very thin black string falls perpendicularly and smoothly into a very large black hole.
Their radius are denoted by $r_{bs}$ and $r_{hs}$ respectively.
Both the black string and the black hole rotate with the same angular velocity $\alpha$ and the axes of the rotation is the direction of the falling  black string flow.
Since we have $r_{bh}\gg r_{bs}$, given that the surface gravity is inverse ratio to the radius, we know that the surface gravity of black string is much larger than the surface gravity of black hole.
When the string is falling into the black hole horizon without any external interference, we can anticipate that the two horizons fuse smoothly into each other.
By this construction, the black hole can accumulate mass from the falling string and, as the consequence, the black hole must grow in size irreversibly.

However, when we take the limit $r_{bh}\to \infty$ and hold $r_{bs}$ fixed, and then just pay attention on events that happen in the intersected region we can get rid of the effect of the growing of the black hole.
The black hole horizon then becomes a Rindler-like infinite horizon because of the rotation.
If the angular velocity is zero, the horizon is exactly the Rindler horizon.
By changing to the rest frame of the falling black string, the acceleration horizon vanishes. Then we have a configuration of a stationary black string.
Alternatively, if we take a stationary black string and study it from a frame that accelerates along the direction of the string, what we observe is that a string free falls into an acceleration horizon.
We will construct the event horizon for such acceleration observers and study how the rotation will affect the result, and then make a comparison with the non-rotation result.

The arrangement of this paper are: in Section 2 we will construct a Kerr black string flow and study the properties of the construction.
In Section 3 we will study the black string flow and its equilibrium condition, and then extend to a more general model. The last section is our conclusion and outlook.
\section{Construction of Kerr black string flow}
Configurations of higher dimensional rotating and charged black holes obtained in string theory have been studied a lot on a micro level \cite{23,24,25}.
Unlike these constructions above, we follow \cite{1} to build up a new metric.
\subsection{Horizon of black string flow}
In article \cite{1} the authors give their metric as follows
\begin{equation}
	ds^2=-f\left(r\right)dt^2+dz^2+\frac{dr^2}{f\left(r\right)}+r^2 d\Omega_{n+1},
\end{equation}
it is a black string flow without rotation in $D=n$ spacetime dimensions.
Generally we can generalize metric (2.1) to a new expression with rotation
\begin{eqnarray}
	ds^2 &=& -\left(1-\frac{2Mr}{\rho ^2}\right)dt^2+\frac{\rho ^2}{\Delta}dr^2+\rho ^2 d\theta ^2 +[\left(r^2+\alpha ^2\right)sin^2 \theta +\frac{2Mr\alpha ^2 sin^4 \theta}{\rho ^2}]d\phi ^2 \nonumber\\
	&-&\frac{4Mr\alpha sin^2 \theta}{\rho ^2}dtd\phi + Fdz^2,
\end{eqnarray}
where
\begin{eqnarray}
	\rho ^2 = & r^2 + \alpha ^2 cos^2 \theta, \\
	\Delta = & r^2 - 2Mr + \alpha ^2,
\end{eqnarray}
and F is the metric function that should have a positive sign outside the black string horizon and be finite at $r \to \infty$ when $\alpha$ is fixed.
The metric (2.2) is in the rest frame of the free-falling and rotating black string in D=5 dimensions.
When the black string is missing $\left( M=0 \right)$, we have the null surfaces $\left(t=z+t_0\right)$ that are the acceleration horizons with different constant $t_0$.
One condition we need to guarantee is that when it is far from the string and at the same time the rotation can be neglected, the null rays can change to the conventional Rindler horizon, i.e.,
\begin{equation}
	\frac{\dot{t}}{\dot{z}}\to 1 ~ ~ ~ and ~ ~ ~ \dot{r} \to 0 ~ ~ ~ for ~ ~ ~  r \to \infty,
\end{equation}
where the dot indicates the derivative of an affine parameter $\lambda$.
By transforming the metric into Boyer-Lindquist coordinate, we have
\begin{equation}
	ds^2=-\frac{\Delta}{\rho ^2}dT^2 +\frac{\rho ^2}{\Delta}dR^2 +\rho ^2 d\Omega +FdZ^2,
\end{equation}
with
\begin{equation}
	d\Omega =d \Theta ^2 + sin^2 \Theta d\Phi.
\end{equation}
The relations between new and old metric are
\begin{eqnarray}
	dT &=& dt-\alpha sin^2 \theta d\phi, \\
	dR &=& dr, \\
	d\Phi &=& \frac{\left(r^2 +\alpha ^2\right)d\phi-\alpha dt}{\rho ^2},\\
	d\Theta &=& d\theta, \\
	dZ &=& dz .
\end{eqnarray}
Since the affine parameter $\lambda$ is also valid in new coordinate system, we have the null path
\begin{equation}
	-\frac{\Delta}{\rho ^2}\dot{T}^2 +\frac{\rho ^2}{\Delta}\dot{R}^2+F \dot{Z}^2 =0,
\end{equation}
and
\begin{equation}
	\dot{T}=\frac{\epsilon \rho ^2}{\Delta}, ~ ~ ~ \dot{Z}=\frac{p}{F}.
\end{equation}
In equation (2.14), $\epsilon$ and $p$ are two integration constants because of the isometries generated by $\partial _T$ and $\partial _Z$. The details are given in appendix A.
From equations (2.5), (2.8) and (2.12) we know the relations between $\epsilon$ and $p$
\begin{eqnarray}
	\frac{dT}{d\lambda}&=&\frac{dt}{d\lambda}-\alpha sin^2\theta \frac{d\phi}{d\lambda},\\
	\frac{dZ}{d\lambda}&=&\frac{dz}{d\lambda},
\end{eqnarray}
and because
\begin{equation}
	\frac{dt}{d \lambda}=\frac{dz}{d\lambda}~ ~ ~ when~ ~ ~  r\to \infty,
\end{equation}
we get
\begin{equation}
	\epsilon + \alpha sin^2\theta \frac{d\phi}{d\lambda}=p.
\end{equation}
We can neglect the contributions of F and $\frac{\rho^2}{\Delta}$ because when $r \to \infty$, both of them approximate to 1.
\subsection{Null hypersurface and the analysis}
Now from the equations (2.14) and (2.18), we have the representation of $\dot{Z}$
\begin{equation}
	\dot{Z}=\frac{\Delta \dot{T}+\rho^2 A_\theta}{F\rho^2},
\end{equation}
with
\begin{equation}
	A_\theta=\alpha sin^2 \theta \frac{d\phi}{d\lambda}.
\end{equation}
Using equations (2.13) and (2.19), we have
\begin{equation}
	 \dot{R}=\pm\sqrt{\left(\frac{\Delta}{\rho^2}\right)^2\dot{T}^2-\frac{\Delta}{F\rho^2}\left(A_\theta+\frac{\Delta}{\rho^2}\dot{T}\right)^2}.
\end{equation}
Because both of the $\dot{Z}$ and $\dot{R}$ can be represented by $\dot{T}$, we can deduce an equation that consists of $\dot{Z}$, $\dot{T}$ and $\dot{R}$,
\begin{equation}
	\dot{T}=\dot{Z}\mp\frac{\Delta \dot{T} -F\rho^2\dot{T} +\rho^2 A_\theta}{F \sqrt{\Delta^2 \dot{T}^2 -\frac{\Delta\left(\Delta \dot{T} +\rho^2 A_\theta\right)^2}{F\rho ^2 }}}\dot{R}.
\end{equation}
Equation (2.22) can be proved by putting equations (2.19) and (2.21) into equation (2.22).
Then the null hypersurfaces ruled by the one-form equation are
\begin{equation}
	dT=dZ\mp\frac{\Delta dT -F\rho^2dT +\rho^2 \hat{A}_\theta}{F \sqrt{\Delta^2 dT^2 -\frac{\Delta\left(\Delta dT +\rho^2\hat{A}_\theta\right)^2}{F\rho ^2 }}}dR,
\end{equation}
where $\hat{A}_\theta=\alpha sin^2 \theta d\phi.$
Back to the old coordinate system we get
\begin{equation}
	dt-\alpha sin^2 \theta d\phi =dz\mp\frac{\left(dt-\alpha sin^2\theta d\phi \right)\Delta -F\left(dt-\alpha sin^2 \theta d\phi\right)\rho^2 +\rho^2 \hat{A}_\theta }{F\sqrt{\left(dt-\alpha sin^2\theta d\phi\right)^2 \Delta^2-\frac{\Delta\left(\left(dt-\alpha sin^2 \theta d\phi\right)\Delta+\rho^2 \hat{A}_\theta\right)^2 }{F\rho^2 } } }dr.
\end{equation}
When $\alpha =0$, Kerr metric will return to Schwarzschlid metric. By setting F=1 and $\frac{\Delta}{\rho^2}=f$ for convenience, the null hypersurfaces become
\begin{equation}
	dt=dz\pm\frac{\sqrt{1-f}}{f}dr.
\end{equation}
These are the null hypersurfaces ruled by geodesics in \cite{1}.

Because of the missing symmetry, equations (2.23) and (2.24) both have extremely complicated resolutions.
For convenience and no losing generality, we first set $F=1$ and $M=1$, then just study the event horizon at $\theta=0$.
In this way we get the new one-form equation
\begin{equation}
	dT=dZ\mp\frac{\left(\Delta dT -\rho^2 dT\right) }{\sqrt{\Delta^2 dT^2-\Delta^3 \rho^{-2}dT^2 } }dR.
\end{equation}
Benefiting from $\theta =0$ the old coordinate system is equal to new one and as a consequence we get
\begin{equation}
	dt=dz\mp\frac{\left(r^2+\alpha^2\right)^{\frac{1}{2}}\sqrt{2r}}{r^2-2r+\alpha^2 }dr.
\end{equation}
Then we have
\begin{equation}
	t=z+t_0\mp\int \frac{\left(r^2+\alpha^2\right)^{\frac{1}{2}}\sqrt{2r}}{r^2-2r+\alpha^2 }dr,
\end{equation}
each value of $t_0$ results in a corresponding null hypersurface that ends at different value in the null coordinate in future null infinity.
Clearly any of them can be regarded as our event horizon.
Here, for convenience,  we denote the null hypersurface with $t_0=0$ as the horizon.
This is not a stationary horizon: the action of $\partial_t$ can change $t_0$, therefore it does not map a hypersurface onto itself but onto another one instead.
The explicit form of the integral result of $r$ in (2.28) is given it in the appendix B.
The surfaces $H_f$ are plotted in figure 1.
\begin{figure}
	\centering
	\includegraphics[width=2in ,clip]{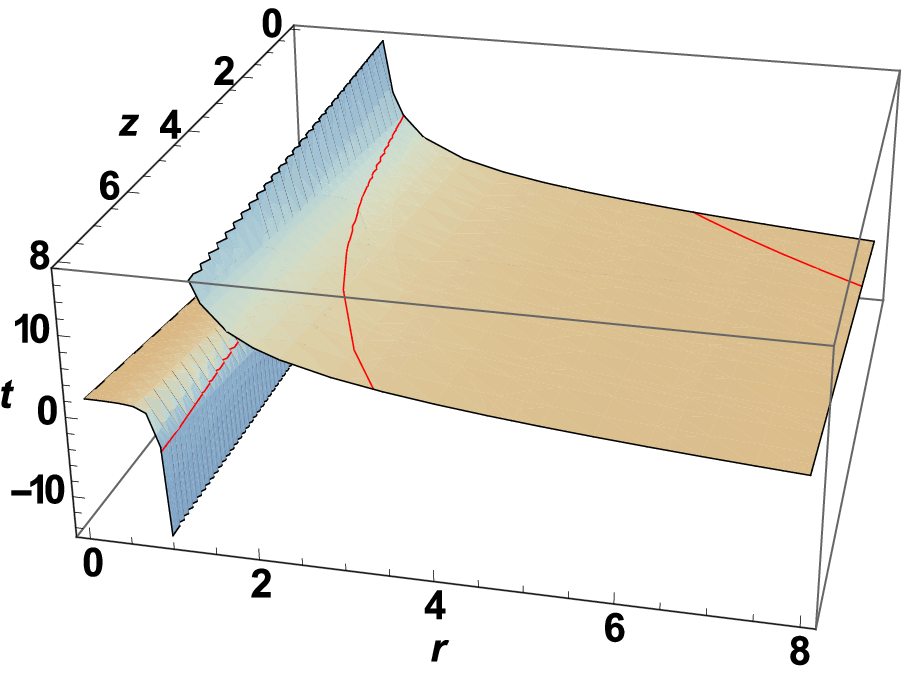}	
	\includegraphics[width=2in ,clip]{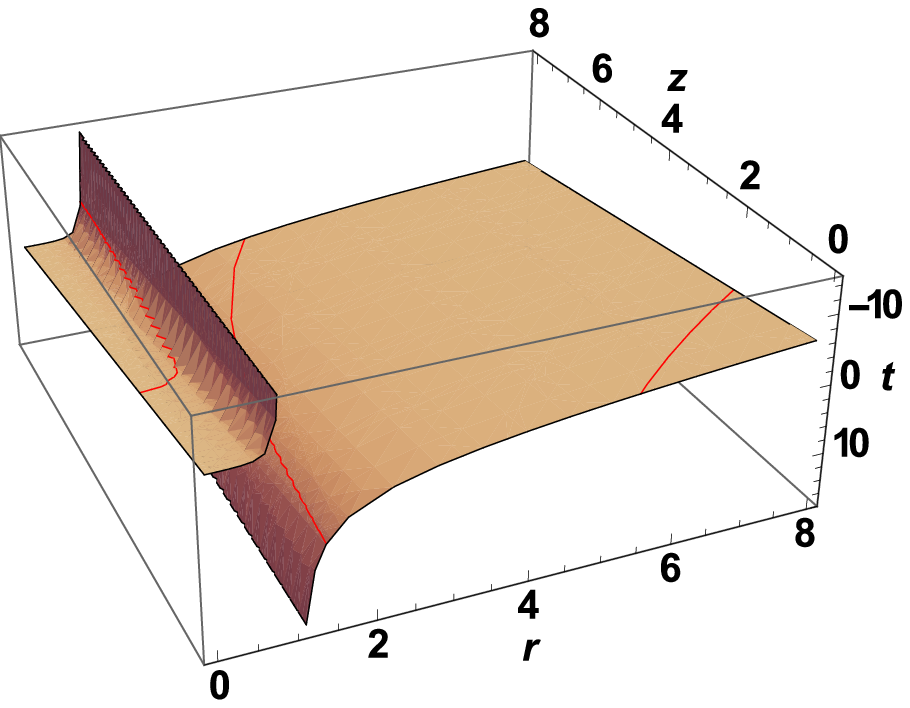}
	\hfill
	\caption{\label{fig:1}Null hypersurfaces $H_f$ of the black string flow are plotted from equation (2.28) in (t, r, z) space for $F=M=\alpha =1$ and $\theta=0$. The left one is the out-going hypersurface which means we choose the positive sign in equation (2.28). The right one is the in-going hypersurface. The red curves are the constant-t section.}
\end{figure}

In these pictures, $M=\alpha$ indicates that they are extreme cases, the inner event horizon and outer event horizon meet at $r=1$.
A constant imaginary part occurs when $r<1$, we use the magnitude of this complex number to get these figures.
\subsection{Null geodesic congruence}
Without losing generality, we just consider the negative term in equation (2.23) which corresponds to the outgoing hypersurface.
Benefiting from the freedom to scale $\lambda$, we can set $\epsilon =p-A_\theta=1$.
Then from the equation (2.14), we have
\begin{equation}
	\dot{T}=\frac{\rho ^2}{\Delta}, ~ ~ ~ \dot{Z}=\frac{1+A_\theta}{F}.
\end{equation}
According to equation (2.22), we have
\begin{equation}
	\frac{\rho^2}{\Delta}=\frac{1+A_\theta}{F}-\frac{\rho^2 -F\frac{\rho^4}{\Delta} +\rho^2 A_\theta}{F\sqrt{\Delta}\rho \sqrt{\frac{\rho^2 }{\Delta}-\frac{\left(1 + A_\theta\right)^2}{F }}}\dot{R}.
\end{equation}
This gives us
\begin{equation}
	\dot{R}=\sqrt{\frac{-\Delta+F\rho^2-2\Delta A_\theta-\Delta A^2 _\theta}{F\rho^2}}.
\end{equation}
Because of $dR=dr$, equation (2.31) can be rewritten in $\left( t, r, z \right)$ coordinate system easily
\begin{equation}
	\frac{dr}{d\lambda}=\sqrt{\frac{C_2r^2+C_1r+C_0}{F\left(r^2+\alpha^2cos^2\theta\right)} },
\end{equation}
with
\begin{eqnarray}
	C_0 &=&\alpha^2\left(cos^2\theta F-\left(A_\theta+1\right)^2\right),\nonumber\\
	C_1 &=& 2M\left(1+A_\theta\right)^2,\nonumber \\
	C_2 &=&F-1-2A_\theta-A^2 _\theta.\nonumber
\end{eqnarray}
Again we consider the situation in figure 1.
By setting $\theta=0$ and $ F=M=\alpha=1$, we get
\begin{equation}
	\frac{dr}{d\lambda}=\sqrt{\frac{2r}{r^2+1} },
\end{equation}
The function $\lambda\left(r\right)$ is plotted in figure 2.
More specific calculations see appendix C.
\begin{figure}
	\centering
	\includegraphics[width=2in ,clip]{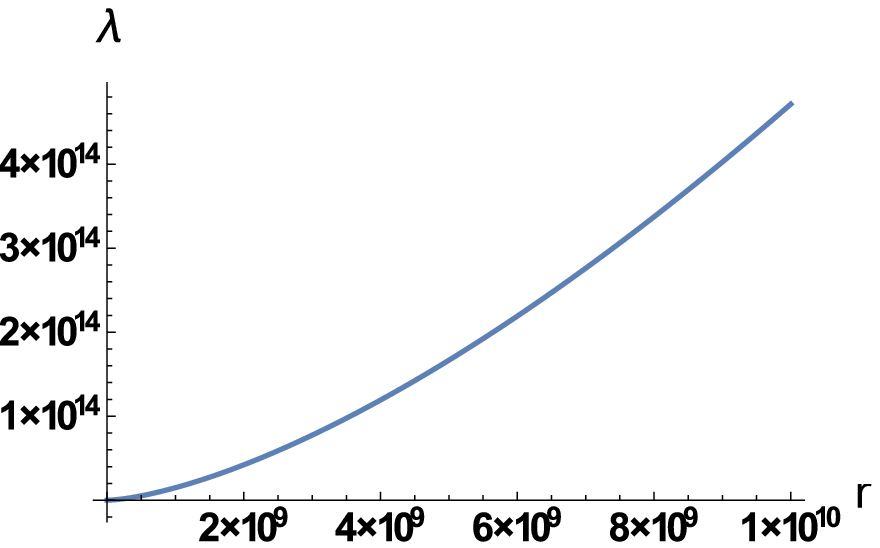}
	\includegraphics[width=2in ,clip]{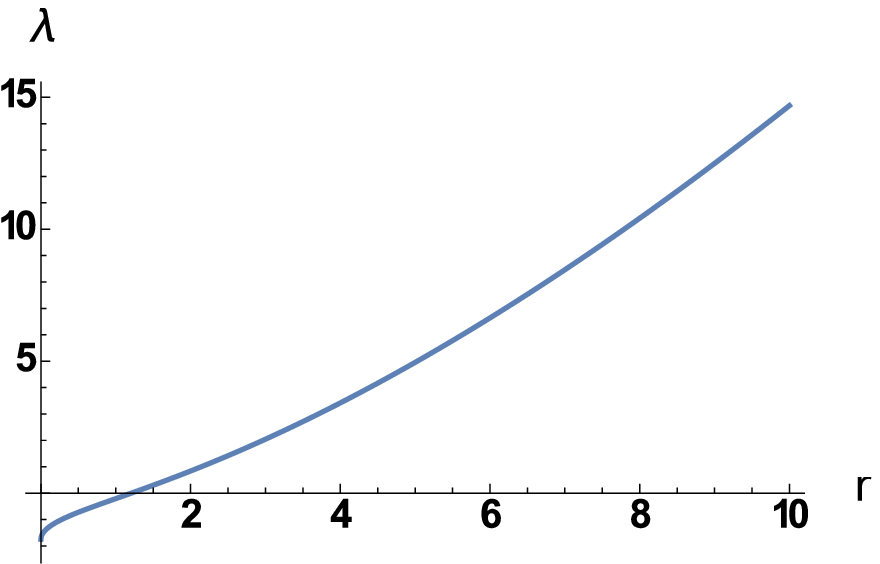}
	\hfill
	\caption{\label{fig:2}The behavior of function $\lambda \left(r\right)$. The picture on the left depicts the function when r is large. And on the right we show the behavior of the function when r is close to 0, and one should notice that 0 is a singularity. When $r \to 0 ,$ we get $\lambda \to \infty$. And as $\lambda$ grows, r moves toward to $\infty$.}
\end{figure}
\subsection{Event horizon and ergosurface}
In Kerr black hole, the region between the outer event horizon and the Killing horizons is known as the ergosphere.
Inside the ergosphere, any objects have to rotate with the black hole, but at the same time they are free to move toward or away from the event horizon.
When we consider the Kerr black string flow, the same thing happens on hypersurface $H_f$.

According to equation (2.33) and figure 2, we find that all null rays on $H_f$ must go towards $r \to \infty$ as the $\lambda$ grows.
This means that any timelike trajectory that remains within the bounded values of r will cross $H_f$ at last.
As a consequence, any observers that remain within a finite range of the black string will fall across $H_f$ eventually.
As the same as the story in Kerr black hole, any observers inside this surface cannot remain static but are dragged along with the string, so this is a particular ergoshpere that the Kerr string flow can distinguish from an ordinary Kerr black hole.

In our construction the black string is accelerating, and then the ergoregion grows.
For any observes who wants to avoid to fall across this $H_f$, an acceleration in the z direction is not enough, they also need to move out towards $r \to \infty$.
Even though the hypersurface $H_f$ we consider here is a special case, it is straightforward to extend this analysis to any $\theta$.
\section{Out-of-equilibrium flow}
Let us consider the vector
\begin{eqnarray}
	\frac{d}{d\lambda} &=& \dot{T}\frac{\partial}{\partial T}+\dot{Z}\frac{\partial}{\partial Z}+\dot{R}\frac{\partial}{\partial R}\\
	&=& \frac{\rho ^2}{\Delta}\frac{\partial}{\partial T}+\frac{\left(1+A_\theta\right)}{F}\frac{\partial}{\partial Z}+\sqrt{\frac{-\Delta+F\rho^2-2\Delta A_\theta-\Delta A^2 _\theta}{F\rho^2}}\frac{\partial}{\partial R}.\nonumber
\end{eqnarray}
This is an affine generator of the null geodesic congruence.
It is convenient to consider the following non-affine generator\footnote{According to the definition in \cite{carroll}, any parameter $\lambda$, which relates to the proper time $\tau$ in this way: $\lambda=a\tau+b$ where $a$ and $b$ are constants, is an affine parameter. It is obvious that $\frac{1}{2}\frac{\Delta}{\rho^2}$ is not a constant, so parameter $l$ is not an affine parameter and the generator of $l$ is not an affine generator, too.} of the future horizon
\begin{eqnarray}
	l &=& \frac{1}{2} \frac{\Delta}{\rho^2}\frac{d}{d\lambda} \\
	&=& \frac{1}{2}\left(\frac{\partial}{\partial T}+\frac{\Delta\left(1+A_\theta\right)}{\rho^2F}\frac{\partial}{\partial Z}+\frac{\Delta}{\rho^2}\sqrt{\frac{-\Delta+F\rho^2-2\Delta A_\theta-\Delta A^2 _\theta}{F\rho^2}}\frac{\partial}{\partial R}\right).\nonumber
\end{eqnarray}
The reason we normalize the generator into equation (3.2) is these generators can recover to the black string horizon and the acceleration horizon near and far from the black string respectively.

The surface gravity $\kappa_{\left(l\right)}$ of $l$ is defined in this form
\begin{equation}
	\nabla_l l=\kappa_{\left(l\right)}l.
\end{equation}
Given that $\lambda$ is an affine parameter, we have
\begin{eqnarray}
	\kappa_{\left(l\right)}&=&\frac{1}{2}\frac{d}{d\lambda}\left[\frac{\Delta}{\rho^2}\right]\nonumber\\
	&=&\frac{-M\alpha^2 cos^2 \theta-\alpha^2r+r\alpha^2cos^2\theta+Mr^2}{\left(\alpha^2cos^2\theta+r^2\right)^2}\frac{dr}{d\lambda}.
\end{eqnarray}
The specific calculations are showed in Appendix D.
According to equation (2.32), we have
\begin{equation}
	\kappa_{\left(l\right)}=\left(\frac{-M\alpha^2 cos^2 \theta-\alpha^2r+r\alpha^2cos^2\theta+Mr^2}{\left(\alpha^2cos^2\theta+r^2\right)^2}\right)\sqrt{\frac{C_2r^2+C_1r+C_0}{F\left(r^2+\alpha^2cos^2\theta\right)} }.
\end{equation}
This surface gravity diminishes monotonically to 0 when $r$ is large enough.
And when $\alpha =0$ the Kerr black string flow returns to Schwarzschlid black string flow. By setting $F=1$ we get the surface gravity
\begin{equation}
	\kappa_{\left(l\right)}=\frac{1}{2r}\left(\frac{2M}{r}\right)^{\frac{3}{2}},
\end{equation}
which coincides with the result in \cite{1}.

There is a chance that by setting a specific $F$ we can make the surface gravity $\kappa=0$.
Namely, putting concrete coefficients $C_i~\left(i=0,1,2\right)$ into equation (3.5), we have
\begin{equation}
	F=\frac{\left(-2Mr+r^2+\alpha^2\right)\left(1+2A_\theta+A^2 _\theta\right)}{r^2+\alpha^2 cos^2\theta}.
\end{equation}

Now we set $M=1$ and $\frac{d\phi}{d\lambda}=\alpha$ to concretely give different expressions of equation (3.7).
We plot this function in figure 3 with four different conditions ($\alpha=$1, 0.75, 0.5, 0,25 respectively).

\begin{figure}
	\centering
	\includegraphics[width=1.6in ,clip]{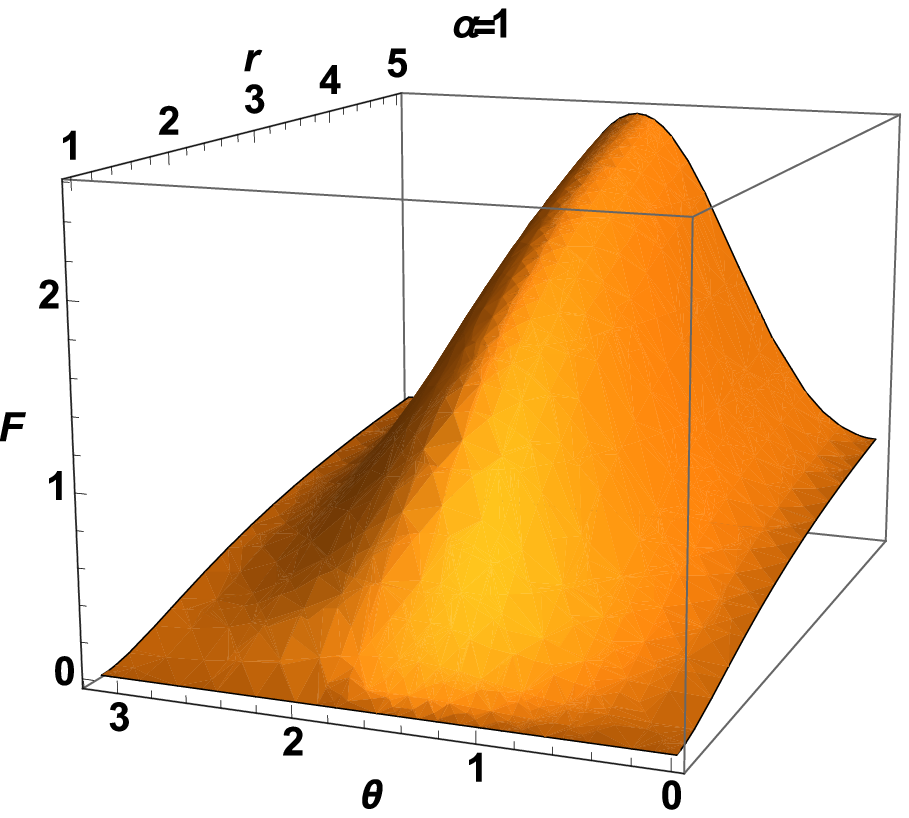}
	\includegraphics[width=1.6in ,clip]{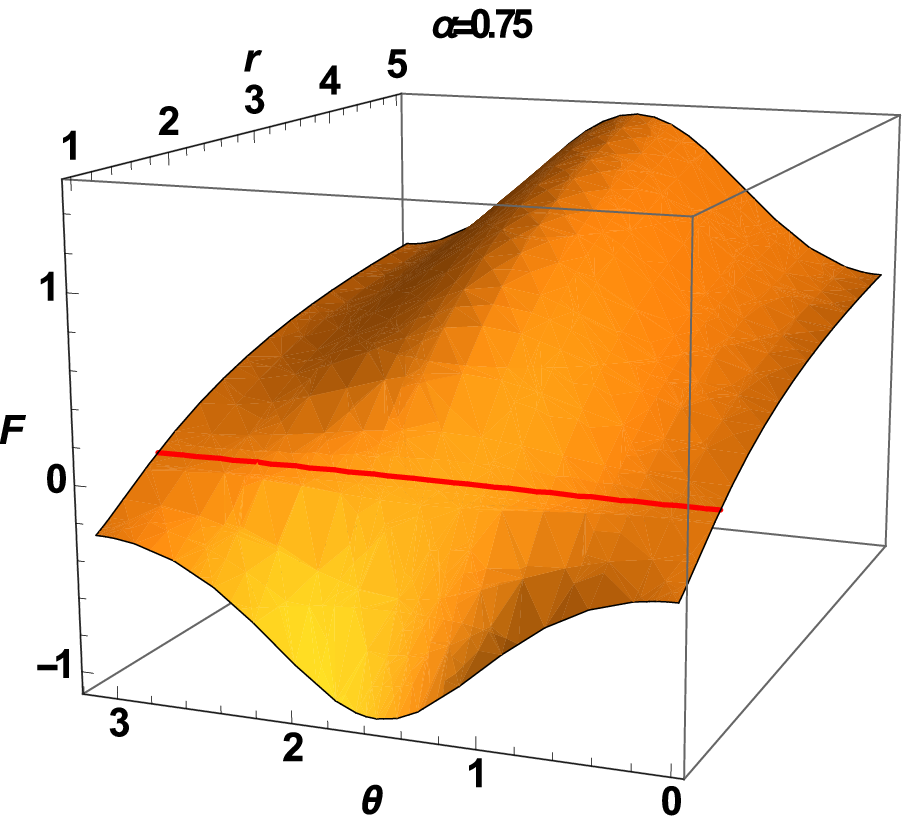}
	\\
	\includegraphics[width=1.6in ,clip]{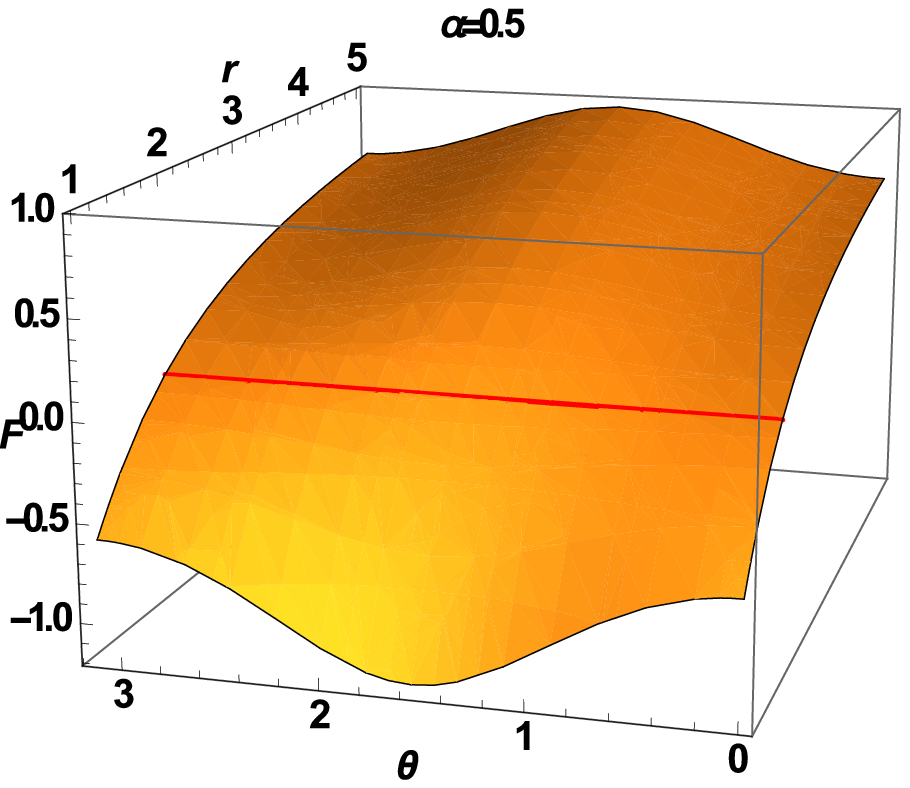}
	\includegraphics[width=1.6in ,clip]{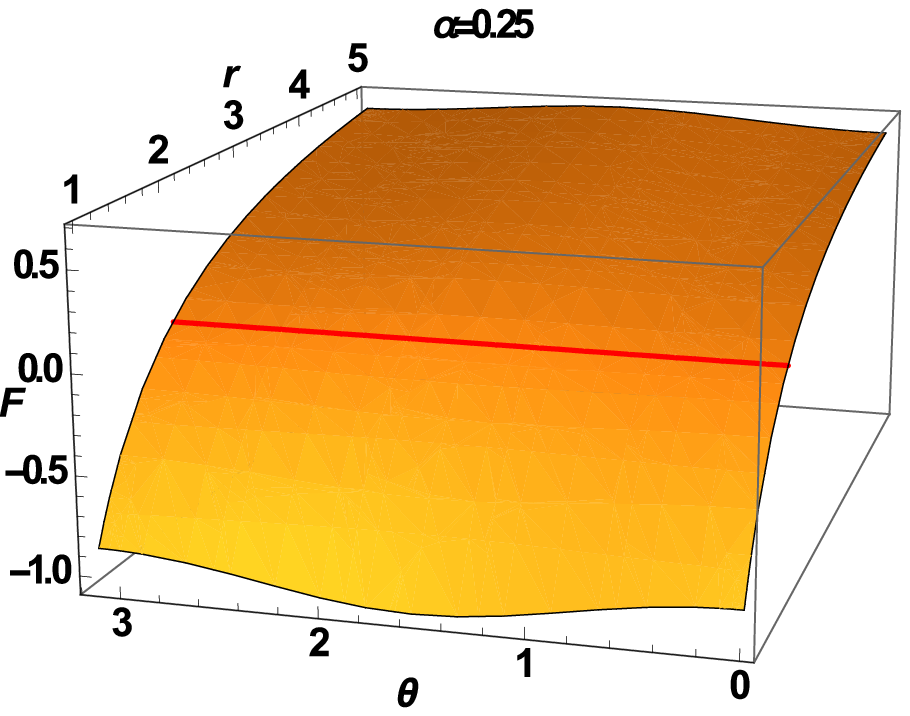}
	\hfill
	\caption{\label{fig:3} In these pictures we show the function $F\left(r,\alpha,\theta\right)$ with four different $\alpha$. The first one is the extreme condition that $M=\alpha=1$, the event horizon locates at $r=1$. Other three pictures are ordinary conditions the red lines denote their outer horizons $r_H$, when $r>r_H$, the function $F$ is positive.}
\end{figure}

Our previous analysis about metric (2.6) can be extended to a more general form, for example, the charged and rotating black string.
The metric can be written in this form
\begin{equation}
	ds^2=-\mathcal{T}dT^2+\mathcal{Z}dZ^2+\frac{dR^2}{\mathcal{R}}+R^2d\Omega ,
\end{equation}
where $\mathcal{T},\mathcal{Z}$ and $ \mathcal{R}$ are general functions of angular momentum J, charge Q and the space coordinates $\left( r, \theta, \phi \right)$. The metric (3.8) is represented in Boyer-Lindquist coordinate by applying equations (2.7)-(2.12).
Here we restrict ourselves to $\mathcal{T}<\mathcal{Z}$.\footnote{When $\mathcal{T}=\mathcal{Z}$ the string worldsheet is Lorentz-invariant. A more circumstantial analysis sees \cite{1}.}
Similar to the deduction and calculation of equation (2.23), we  achieve the null hypersurface
\begin{equation}
	 dT=dZ+\frac{\mathcal{Z}dT-\mathcal{T}dT-\hat{A}_\theta}{\mathcal{Z}\sqrt{\mathcal{R}}\sqrt{\mathcal{T}dT^2-\frac{\left(\mathcal{T}dT+\hat{A}_\theta\right)^2}{\mathcal{Z}}}}dR,
\end{equation}
where $\hat{A}_\theta = \alpha sin^2 \theta d\phi$.

Similar to the deductions of equations (2.29) and (2.31), the congruences in terms of the affine parameter $\lambda$ are
\begin{eqnarray}
	dZ&=&\left(1+\hat{A}_\theta\right)\mathcal{Z}^{-1}d\lambda+d\xi,\\
	 dR&=&\sqrt{\mathcal{R}}\left(\frac{1}{\mathcal{T}}-\frac{\left(1+\hat{A}_\theta\right)^2}{\mathcal{Z}}\right)^{1/2}d\lambda ,
\end{eqnarray}
here the letter $\xi$ is to label the different null rays.
Then, following the beginning of section 3, we can give the non-affine null geodesic generator
\begin{eqnarray}
	l &=&\frac{1}{2}\mathcal{T}^{-1}\frac{d}{d\lambda}\\
	&=& \frac{1}{2}\left(\frac{\partial}{\partial T}+\frac{\mathcal{T}\left(1+A_\theta\right)}{\mathcal{Z}}\frac{\partial}{\partial Z}+\sqrt{\mathcal{R}}\sqrt{\frac{1}{\mathcal{T}}-\frac{\left(1+A_\theta\right)^2}{\mathcal{Z}}}\frac{\partial}{\partial R}\right)\nonumber,
\end{eqnarray}
where  $A_\theta=\alpha sin^2 \frac{d\phi}{d\lambda}$.
And similar to appendix D, because $\lambda$ is an affine parameter, we have the surface gravity $\kappa_{\left(l\right)}$
\begin{eqnarray}
	\kappa_{\left(l\right)}&=&\frac{1}{2}\frac{d}{d\lambda}\left[\mathcal{T}\right]\nonumber\\
	&=&\frac{1}{2}\frac{dR}{d\lambda}\frac{\partial \mathcal{T}}{\partial R}\\
	 &=&\frac{1}{2}\sqrt{\mathcal{R}}\sqrt{\frac{1}{\mathcal{T}}-\frac{\left(1+A_\theta\right)^2}{\mathcal{Z}}}\frac{\partial \mathcal{T}}{\partial R}.\nonumber
\end{eqnarray}
When the rotation decreases to 0, we get the ordinary result
\begin{equation}
	 \kappa_{\left(l\right)}=\frac{1}{2}\sqrt{\left(\mathcal{Z}-\mathcal{T}\right)\frac{\mathcal{R}}{\mathcal{T}\mathcal{Z}}}\frac{\partial \mathcal{T}}{\partial r}.
\end{equation}
Again this is the result in the static spacetime.
In equation (3.13), we can get an equilibrium condition 
\begin{equation}
	\mathcal{Z}=\left(1+A_\theta\right)^2\mathcal{T}.
\end{equation}
This extreme limit will happen when we consider the black strings with basic string charge.
The charged string flow describes a string with a specific excitation down to a large black hole.
Even if the string charge can decrease the temperature of the black string, only at equation (3.15) can it be in thermal equilibrium.
Above the extreme situation, the string excitation is much hotter than that of the black hole, and the system seems to have a different behavior from the configuration which the string excitation is in thermal equilibrium with a finite temperature horizon (a worldsheet approach \cite{a,b,c,d} or a blackfold approach \cite{e,f}).
In this formation, when the black string does not attain to extreme, it will dump energy to the black hole.
\section{Conclusion and outlook}
In this paper, we generally give and analyze a model of the rotating black string flow in dimension D=5, and extend this solution to a charged and rotating black sting flow, and study their equilibrium conditions. In these constructions, the smooth intersection between the black hole and black string is in the axisymmetric spacetime, which has directly practical physics picture.
The system we study here requires that the string is very thin comparing with the black hole and the whole progress is free from any external force interference.

When studying the instability of the black string in the late time evolution, a similar construction has been found \cite{8}.
The microscopic construction of the place where the black string and black hole intersect is very interesting.
Another interesting field is the similarity between the black funnels and the black string flow \cite{1,2,7,9,10,21,20,13,15,16}.
Furthermore, there should be some relations between the rotating black string flow and some kinds of black funnels.
Using this method we can extend to other situations, for example, a free falling string that is not rotated falls into a rotating black hole.
Under this construction we may expect the rotation velocity of the black hole decreases to zero after the string pour enough mass to this system. Therefore, a lot of relevant works can be done.
\section*{Acknowledgments}
We gratefully acknowledge useful discussions with Dr. Ding-Fang Zeng, Dr. Jian-Feng Wu and all the others in my institute. The work is supported by National Natural Science Foundation of China (No. 11275017 and No. 11173028).

\appendix
\section{Calculation in horizon of black string flow}
Details of deducing equation (2.14): first considering this equation
\begin{equation}
	I\left(x^A\right)=\int g_{AB} \frac{dx^A}{d\lambda}\frac{dx^B}{d\lambda}d\lambda,
\end{equation}
where A, B=(0,1, 2, 3, 4), then we have
\begin{equation}
	\partial_T I=\epsilon=\partial_T\int g_{AB} \frac{dx^A}{d\lambda}\frac{dx^B}{d\lambda}d\lambda,
\end{equation}
and then
\begin{eqnarray}
	\epsilon &=& \partial_T \int g_{AB}\left(\frac{d}{d\lambda}\left(x^A\frac{dx^B}{d\lambda}\right)-x^A\frac{d^2x^B}{d\lambda^2}\right)d\lambda\\
	\epsilon &=& -\int g_{00}\partial_T x^0\frac{d^2x^0}{d\lambda^2}d\lambda\nonumber\\
	\epsilon &=& \frac{\Delta}{\rho^2}\frac{dT}{d\lambda}\nonumber\\
	\frac{dT}{d\lambda}&=&\frac{\epsilon \rho^2}{\Delta}\nonumber.
\end{eqnarray}
We leave the boundary term away here as in field theory.
By this method we can get $\frac{dZ}{d\lambda}=\frac{p}{F}$, too.
\section{Explicit integration in null hypersurface}
From (2.28) and set $\alpha =1$, we have
\begin{eqnarray}
	t-t_0 	&=& z+\sqrt{2} \int \frac{\sqrt{r^3+r}}{\left(r-1\right)^2} dr\nonumber\\
	&=& \sqrt{2} \sqrt{r+r^3}\Big[\frac{1}{1-r}+\frac{1}{2\left(r+r^3\right)}\Big[6+6r^2 \nonumber\\
	&-&\left(3-3i\right)\sqrt{2}\sqrt{1+\frac{1}{r^2}}r^{\frac{3}{2}}EllipticE\left(i Arcsinh\left(\frac{1+i}{\sqrt{2r}}\right),-1\right) \nonumber\\
	&+&\left(2-4i\right)\sqrt{2}\sqrt{1+\frac{1}{r^2}}r^{\frac{3}{2}}EllipticF\left(i Arcsinh\left(\frac{1+i}{\sqrt{2r}}\right),-1\right) \nonumber\\
	&+&\left(4+4i\right)\sqrt{2}\sqrt{1+\frac{1}{r^2}}r^{\frac{3}{2}}EllipticPi\left(i,i Arcsinh\left(\frac{1+i}{\sqrt{2r}}\right),-1\right)\Big]\Big].\nonumber
\end{eqnarray}
This result is given by Mathematica 7.0 and will give a constant imaginary part when $r<1$.
Our explanation is when any observers come across the event horizon they will travel in a region where time coordinate and space coordinate exchange their symbols.
In order to denote this effect when they travel into the inner horizon, a constant imaginary part must remain.
\section{Calculation in null geodesic congruence}
From (2.33) we can get
\begin{eqnarray}
	\lambda &=& \int \sqrt{\frac{r^2+1}{2r}} dr,
\end{eqnarray}
then we get
\begin{equation}
	\lambda \left(r\right)=\frac{\sqrt{2}}{3\sqrt{r}}\sqrt{\frac{1}{r}+r}\left(r^{\frac{3}{2}}+\frac{2\left(-1\right)^{\frac{1}{4}}EllipticF\left(iArcsinh\left(\frac{\left(-1\right)^{\frac{1}{4}}}{\sqrt{r}} \right),-1\right)}{\sqrt{1+\frac{1}{r^2}}}\right),
\end{equation}
this is the inverse function of $r\left(\lambda\right)$.
Clearly the equation (2.33) obeys the limitation $\dot{r} \to 0$ when $r\to \infty$ as we mentioned in equation (2.5).
The metric function F is a function of  r, $\alpha$ and $\theta$, so that the equation (2.32) satisfies the limitation in equation (2.5), too.
\section{Calculation in out-of-equilibrium flow}
Using equation (3.3), we have
\begin{equation}
	 \frac{1}{2}\frac{\Delta}{\rho^2}\nabla_{\frac{d}{d\lambda}}\frac{1}{2}\frac{\Delta}{\rho^2}\frac{d}{d\lambda}=\kappa_{\left(l\right)} l,
\end{equation}
then
\begin{equation}
	 \frac{1}{4}\frac{\Delta}{\rho^2}\frac{d}{d\lambda}\left[\frac{\Delta}{\rho^2}\right]\frac{d}{d\lambda}+\frac{1}{4}\frac{\Delta^2}{\rho^4}\nabla_{\frac{d}{d\lambda}}\frac{d}{d\lambda}=\kappa_{\left(l\right)} l.
\end{equation}
Because $\lambda$ is an affine parameter, this means that $\nabla_{\frac{d}{d\lambda}}\frac{d}{d\lambda}=0$.
Then we have
\begin{equation}
	\frac{1}{2}\frac{d}{d\lambda}\left[\frac{\Delta}{\rho^2}\right]l=\kappa_{\left(l\right)} l.
\end{equation}
We now can get equation (3.4).

\end{document}